\documentclass[11pt]{article}
\usepackage{array}
\usepackage{amsmath}
\usepackage{amssymb}
\usepackage{graphicx}
\usepackage{amsthm}
\usepackage{algpseudocode}
\usepackage{latexsym}
\newtheorem{theorem}{Theorem}
\newtheorem{cor}{Corollary}

\newtheorem{prop}{Proposition}
\newtheorem{lemma}{Lemma}
\newtheorem{example}{Example}

\newcommand{\beq}{\begin{equation}}
\newcommand{\eeq}{\end{equation}}
\newcommand{\barr}{\left[\begin{array}}
\newcommand{\earr}{\end{array}\right]}

\newcommand{\bpf}{\emph{Proof}\/:}
\newcommand{\epf}{\hfill$\Box$}

\newcommand{\bi}{\begin{itemize}}
\newcommand{\ei}{\end{itemize}}
\newcommand{\bnum}{\begin{enumerate}}
\newcommand{\enum}{\end{enumerate}}
\newcommand{\bc}{\begin{center}}

\begin{document}
\title{Projective cofactor decompositions of Boolean functions and the satisfiability problem}
\author{Madhav Desai\\madhav@ee.iitb.ac.in
\and 
Virendra Sule\\vrs@ee.iitb.ac.in
\and
Department of Electrical Engineering\\Indian Institute of Technology Bombay, India}
\maketitle
\date{}

\begin{abstract}
Given a CNF formula $F$, we present a new algorithm for deciding the satisfiability (SAT) of $F$ and computing all solutions of assignments. The algorithm is based on the concept of \emph{cofactors} known in the literature. This paper is a fallout of the previous work by authors on Boolean satisfiability \cite{sul1, sul2,sude}, however the algorithm is essentially independent of the orthogonal expansion concept over which previous papers were based. The algorithm selects a single concrete cofactor recursively by projecting the search space to the set which satisfies a CNF in the formula. This cofactor is called \emph{projective cofactor}. The advantage of such a computation is that it recursively decomposes the satisfiability problem into independent sub-problems at every selection of a projective cofactor. This leads to a parallel algorithm for deciding satisfiability and computing all solutions of a satisfiable formula.  
\end{abstract}

\section{Introduction}
In this paper we consider the problem of deciding satisfiability of a Boolean equation 
\[
F(x_{1},\ldots,x_{n})=1
\]
and that of computing all satisfying assignments of variables $x_{i}$ in the Boolean algebra $B_{0}=\{0,1\}$. This is a celebrated problem of computer science with vast applications \cite{knut,crah,rude}. We shall denote by $B(n)$ the Boolean algebra of all Boolean functions $f:B^{n}\rightarrow B$ of $n$ variables over a Boolean algebra $B$. For all other notations on Boolean functions we refer to \cite{brow} and to \cite{hand} for satisfiability literature.  

Although this paper is a sequel to previous papers \cite{sul1, sul2, sude} in which the satisfiability of the Boolean equation $F=0$ was also considered, the central issue addressed in these papers was that of representation of Boolean functions $F$ in several variables in the Boole-Shannon expansion form and to express the satisfiability and all solutions of a satisfiable equation in terms of the expansion co-efficient functions with respect to orthonormal functions. In \cite{sude} it was shown that actually the orthonormality itself was not necessary for such an expansion and the expansion formula was vastly generalized. To recapitulate the ideas we shall re-state these results of \cite{sude} in a comprehensive form as follows \cite[theorem 1,2, corollary 1]{sude}.

\begin{theorem}
 Let $G=\{g_1, g_2, \ldots g_m \}$ be a set of non-zero Boolean functions and a Boolean function $f$ be such
that 
\[
f \leq g_1 + g_2 + \ldots + g_m
\]
Then, 
\begin{enumerate}
\item $f$ can be expressed as 
\begin{equation} \label{eq:StdExp}
f \ = \ \alpha_1.g_1 + \alpha_2.g_2 + \ldots + \alpha_m.g_m
\end{equation}
where each $\alpha_i$, $i=1,2,\ldots m$, is a Boolean function, called \emph{cofactor} of $f$ wrt $g_{i}$, 
which can be chosen freely in the interval $[f.g_i,\ f + g_i']$.
\item $f=0$ is consistent iff for some $i$ the system of equations $\alpha_{i}=0$ and $g_{i}=1$ is consistent, (i.e. $\alpha_{i}=0$ has a solution on the set of satisfying assignments of $g_{i}$ for some $i$) for an arbitrary choice of cofactor $\alpha_{i}$ in its interval of existence.
\item All solutions of $f=0$ arise as the union of all solutions of systems $\alpha_{i}=0$, $g_{i}=1$ whenever the later systems are consistent.
\end{enumerate}
\end{theorem}

Hence this expansion of $f$ in terms of $g_{i}$ gives decomposition of the set of all satisfying assignments for $f=0$ in terms of independent problems. Such a decomposition is thus of great importance for computation. In fact as the later result \cite[corollary 2]{sude} shows an algorithm for satisfiability of a CNF formula $F$ and computation of all satisfying solutions does not require actual computation of the cofactors $\alpha_{i}$. The algorithm developed requires generation of independent reduced formulas $f_{i}$ at the satisfying assignments of $g_{i}$. This algorithm can thus be called (using terminology from logic) as the \emph{symantic} decomposition algorithm.

\subsection{Transformational algorithm for satisfiability}
In this paper we also take up the problem of describing all satisfying assignments of a CNF formula $F$
\beq\label{CNF}
F=\prod_{i=1}^{m}C_{i}
\eeq
where $C_{i}$ are clauses, from the point of view of transformation of variables. The central idea of the algorithm proposed in this paper is to transform variables so that they successively map to the satisfying set of the partial clauses. This leads to yet another type of a decomposition of $F$ which gives an algorithm for satisfiability and computation of all solutions. This may be called a \emph{transformational} decomposition. Using this central idea we propose in this paper a new algorithm for satisfiability of CNF formulas.

\section{Properties of cofactors and construction}
In \cite{sude} several properties of cofactors of Boolean functions $f$ relative to a base set $\{g_{i}\}$ were derived. For completeness these are summarized again here along with new properties.
The cofactors $\alpha_i$ in the expansion shown in Eq. (\ref{eq:StdExp}) can be
thought of as kind of quotients and their interval of existence is given in the above theorem. We continue with the notation $\Xi(f,g)$ to denote the members of the set of cofactors of $f$ with respect to $g$. From the basic definition of a cofactor \cite[Definition 1]{sude} it follows that elements of $\Xi(f,g)$ are those functions which are restrictions of $f$ on $g_{ON}$ (i.e. match exactly with $f$ for all points on $g_{ON}$). Algebraically all such cofactors belong to the interval $[fg,f+g']$ and conversely every function in this interval is a cofactor. We call $fg$ the \emph{minimal} cofactor and $f+g'$ as the \emph{maximal} cofactor.

\subsection{Algebra of co-factors}
Some of the additional algebraic properties of cofactors are given below. For any two subsets of Boolean functions $A,B$, we define
\begin{eqnarray*}
A + B  & = & \{ u + v : u \in A, \ v \in B \} \\
A.B    & = & \{ u.v : u \in A, \ v \in B \} \\
A'     & = & \{ u' : u \in A \}
\end{eqnarray*}
\begin{prop}\emph{Let $f$, $g$, $h$ be arbitrary Boolean functions. Then
 \begin{enumerate}
\item $\Xi(f,g) + \Xi(f,h) = \Xi(f,g+h)$.
\item $\Xi(f,g).\Xi(f,h)= \Xi (f,g.h)$ as sets.
\item If $g\leq h$ then $\Xi(f,h)\subset \Xi(f,g)$.
\item $\Xi(f+h,g) \ = \Xi(f,g) + \Xi(h,g)$.
\item $\Xi(f.h,g) \ = \Xi(f,g) . \Xi(h,g)$.
\item $\Xi(f',g)=\Xi(f,g)'$.
\end{enumerate}
} \label{prop:CofactorAlgebra}
\end{prop}

The following identity also holds

\begin{prop}
\[
\Xi(f, g.h) \  = \ \{ u.v | u \in \Xi(f,g), v \in \Xi(u,h) \} 
\] \label{prop:CofactoringWithProduct}
\end{prop}
\bpf
By the definition of $u,v$, we see that
$u.g = f.g$ and $v.h = u.h$.  
Thus, 
\begin{eqnarray*}
(u.v).(g.h) & = &  (u.g).(v.h) \\
	& = &  (f.g).(u.h) \\
	& = &  (u.g).(f.h) \\
	& = &  (f.g).(f.h) \\
        & = &  f.(g.h)  
\end{eqnarray*}
so that $u.v \in \Xi(f,g.h)$.
\epf

The expansion shown in Equation \ref{eq:StdExp} expresses $f$ in terms of cofactors wrt a base set $\{g_{i}\}$. We first show an example for computation of minimal and maximal cofactors.
\begin{example}\emph{
Consider $f=x_{1}'x_{2}+x_{2}x_{3}+x_{1}x_{3}'$ and $C=(x_{1}'+x_{3})$ then
\[
fC=x_{1}'x_{2}+x_{2}x_{3}=x_{2}C
\]
which is the minimal cofactor in $\Xi(f,C)$ while
\[
f+C'=f
\]
hence maximal $\Xi(f,C)$ is $f$ itself. A general cofactor in $\Xi(f,C)$ is
\[
fC+pC'=x_{2}(x_{1}'+x_{3})+p(x_{1}x_{3}')
\]
where $p$ is an arbitrary Boolean function.
}
\end{example}

\section{Projective cofactors}

Suppose that  $g \neq 0, h \neq 1$ are Boolean functions in 
$B_{0}(n)$.  For any Boolean function (say $g$), we use $g_{ON}$
to denote the set of points on which $g$ evaluates to $1$ (the
ON-set of $g$), and
$g_{OFF}$ to be the set of points on which $g$ evaluates to $0$
(the OFF-set of $g$).

A {\it projection} is a map $P:B_{0}^{n}\rightarrow B_{0}^{n}$.
Given $g,h$ such that either $g = h = 1$ or $h \neq 1$, we define
a particular kind of projection $P_{g,h}:B_{0}^{n}\rightarrow B_{0}^{n}$ 
as follows:
\begin{displaymath}
P_{g,h}(x)= \lbrace
\begin{array}{ll}
x & \mbox{ if } x \in g_{ON}\\
y_x \in h_{OFF} & \mbox{ if } \ x \in g_{OFF} 
\end{array}
\end{displaymath}
where $y_x$ is an arbitrary element in $h_{OFF}$
which may be chosen independently for each $x$).
That is, the ON-set of $g$ is left unchanged, but the
OFF-set of $g$ is mapped into the OFF-set of $h$.

We call the collection of such projections as the set of
projections of $g$ into $h$, denoted by ${\cal P}(g,h)$.
One possible choice is to select
a fixed $y \in h_{OFF}$ and to set $y_x = y$ for all $x$.
Thus $P_{g,h}$ projects the entire space onto a subset of $h_{OFF} \bigcup  g_{ON}$. 

The following important properties of such projections will be
used in the sequel.
\begin{lemma} \label{lemma:ProjComp}
Let $g_1$, $g_2$ and $h$ be Boolean functions with $h \neq 1$.
Let $P_1$ be a projection in ${\cal P}_{g_1,h}$ and
$P_2$ be a projection in ${\cal P}_{g_2,h}$. 
Then the compositions $(P_1 \circ P_2)$  
and $(P_2 \circ P_1)$ are both
projections in ${\cal P}_{g_1.g_2, h}$.
\end{lemma}
\bpf
Suppose that $u$ is in the ON-set of $g_1.g_2$.  Then u is in ${g_1}_{ON}$ and
${g_2}_{ON}$, and thus $(P_1 o P_2)(u) = u$.  If $u$ is not in the
ON-set of $g_1.g_2$, then $u$ is either in ${g_2}_{OFF}$ or in ${g_1}_{OFF}$ or in both.
Suppose that $u$ is in ${g_2}_{OFF}$, then $w=P_2(u)$ is in $h_{OFF}$.  
Observe that $w$ may or may not be in ${g_1}_{OFF}$. In
either case, it is easy to check that  $P_1(w)$  will be in $h_{OFF}$.
Thus the composition $(P_1 \circ P2)$ maps $u$ appropriately.
The other cases can be handled similarly.
\epf

Given a projection $P$ and some function $h$, we can then obtain the
function
\[
(h \circ P) (x) \ = \ h(P(x))
\]
where $\circ$ denotes composition.
Now if $P$ is chosen to be in ${\cal P}_{g,h}$, we define 
\[
\zeta[h,g,P] \ = \ (h \circ P)
\]
The function $\zeta[h,g,P]$ depends on $P \in {\cal P}_{g,h}$ which in turn depends on the
elements $y_x$ in $h_{OFF}$ used in determining $P$. Further, every $\zeta$
defined in this manner matches with the restriction of $h$ on $g_{ON}$ and 
every such $\zeta$ is a cofactor in $\Xi(h,g)$ and also belongs to the interval $[hg,h+g']$. We call all such cofactors, the \emph{projective cofactors} of $g$ in $h$. 
(Note that if $g \neq 1$, then such a projective co-factor is defined only if $h \neq 1$). 
The projective cofactors of $g$ in $h$ constitute a set of Boolean functions 
parametrized by the choice of elements $y_x \in h_{OFF}$ for each element
$x \in g_{OFF}$. 
A similar (but not identical) concept of a projective cofactor
has been used by Coudert and Madre to construct the {\em restrict} and {\em constrain} operators 
\cite{ref:CoudertMadre}.

Some properties of projective cofactors are easy to establish.
\begin{lemma} \label{lemma:Cofactors}
Let $f$, $g$, $h_1$ and $h_2$ be Boolean functions with
$f \neq 1$, $h_1 \neq 1$ and $h_2 \neq 1$.  Then
\begin{enumerate}
\item If $f \leq g$, then $\zeta[f,g,P] = f$ for all $P \in {\cal P}_{g,f}$.
\item If $f.g  = 0$, then $\zeta[f,g,P] = 0$ for any projection $P \in {\cal P}_{g,f}$.
\item If $h_1 \leq h_2$, then ${\cal P}_{g, h_2} \ \subset \ {\cal P}_{g, h_1}$.
\item If $h \neq 1$, then $\zeta[h,g,P] \leq h$ for all $P \in {\cal P}_{g,h}$.

\end{enumerate}
Further, for any projection $P$, if $f_1$, $f_2$ are Boolean functions, then
\begin{eqnarray}
\zeta[f_1.f_2, g,P] & = & \zeta [f_1,g,P] . \zeta[f_2,g,P] \label{eq:CofAnd} \\
\zeta[f_1+f_2, g,P] & = & \zeta [f_1,g,P] + \zeta[f_2,g,P] \\
\zeta[f',g,P] & = & \zeta[f,g,P]'
\end{eqnarray}
\end{lemma}

\bpf
The conclusions follow from a routine application of the definition of ${\cal P}_{g,f}$ and
$\zeta[]$.
\epf

The importance of projective co-factors lies in the following result.
\begin{theorem} \label{theorem:projectiveSat}
Let $f,g,h$ be Boolean functions with $f = g.h$ and $h \neq 1$.  Then $f$ is satisfiable if and only if, 
for any projection $P \in {\cal P}_{g,h}$, $\zeta[h,g,P]$ is satisfiable.  Further, the solution
set to $f(z)=1$ and $\zeta[h,g,P](z)=1$ are the same, so that every solution of $f(z)=1$
can be obtained by solving $\zeta[h,g,P](z)=1$.
\end{theorem}

\bpf
Since $h \neq 1$, the set of projections ${\cal P}_{g,h}$ contains at least
one element.

Suppose that $f$ is satisfiable and $f(z)=1$. Then $g(z) = 1$ and $h(z)=1$. Observe that
$z$ belongs to $g_{ON}$. Let $P$ be any projection in ${\cal P}_{g,h}$. Clearly $P(z) = z$ because
$z \in g_{ON}$.  Thus $\zeta[h,g,P](z) = h(P(z))=h(z)= 1$. This proves necessity.  

Conversely, assume that for some $z$ and $P \in {\cal P}_{g,h}$, $\zeta[h,g,P](z)=1$. 
Let $u=P(z)$, so that $h(u) = 1$.
Now consider $z$: if $z$ were to be in $g_{OFF}$ then
by the definition of $P$, 
$u = P(z) \in h_{OFF}$ and thus we have $h(u)=0$, a contradiction.
Thus, $z \in g_{ON}$ and consequently $u=z$.  Thus $g(z).h(z) = f(z) = 1$
and $f$ is satisfiable.

We have shown that $f(z)=1$ if and only if $\zeta[h,g,P](z)=1$ to complete
the proof of the last claim.

\epf

\vspace{1cm}

From this result, we infer the following:
\begin{cor} \label{cor:satassign}
Let $f = g.h$.  Choose an arbitrary projection
$P \in {\cal P}_{g,h}$.  
\begin{enumerate}
\item If $g$ and $\zeta[h,g,P]$ are independently satisfiable then $f$ is satisfiable.  
\item The solution set of  $\zeta[h,g,P](x) = 1$ is the same as the
solution set of $f(x)=1$.
\item If one of $g$, $\zeta[h,g,P]$ is identically $0$,
then $f$ is identically $0$.
\end{enumerate}
\end{cor}

Thus, to check satisfiability of $f=g.h$, it is enough to {\em independently} check the satisfiability of $g$ and $\zeta[h,g,P_{g,h}]$ for any choice of $P_{g,h}$. This leads
to a decomposition technique in which the satisfiability of $f$ can be determined in terms of simpler functions $g$
and $\zeta[h,g,P_{g,h}]$.  In general, we may be asked to check the
satisfiability of a Bpplean function given as $f=h_1.h_2.\ldots h_k$, where
each $h_i \neq 1$ is described by a small formula (a clause for example).
Using Theorem \ref{theorem:projectiveSat}, we can show the following.
\begin{theorem} \label{thm:ProjectiveDecomposition}
Let $f = h_1.h_2. \ldots h_k$, with $h_i \neq 0$ for each $i$. 
For $i=1,2, \ldots k$.  Then
$f$ is satisfiable, if and only if the following
functions are satisfiable:
\begin{eqnarray*}
f_1 & =  & h_1,\  {\rm choose} \ P_1 \in {\cal P}_{f_1,h_2} \\
f_2 & =  & \zeta [h_2, f_1, P_1],\  {\rm choose} \ P_2 \in {\cal P}_{f_2,h_3}\\
f_3 & =  & \zeta [ \zeta [h_3, f_1, P_1], f_2, P_2 ],\ {\rm choose}\ P_3 \in {\cal P}_{f_3,h_4} \\
\ldots  \\
f_k & =  & \zeta [ \zeta [ \ldots \zeta [h_k, f_1,P_1], f_2,P_2] \ldots ], f_{k-1},P_{k-1}]
\end{eqnarray*}
where, for $j=1,2, \ldots (k-1)$, $P_j \in {\cal P}_{f_{j},h_{j+1}}$.
Further, if $f_k(x) = 1$, then $f(x) = 1$. 
\end{theorem}

\bpf
  For $j=1,2,\ldots k$, let $u_j = \Pi_{i=j}^{k} h_i$.  Then
$f = h_1. u_2$ and $u_i = h_i. u_{i+1}$ for $i=1,2,\ldots,k-1$.
By Thereom \ref{theorem:projectiveSat}, since $h_i \neq 1$,
$f$ is satisfiable if and only if $\zeta[u_2,h_1,P_1]$ is satisfiable.
But $\zeta [u_2, h_1, P_1] = \zeta[h_2,h_1,P_1]. \zeta[u_3,h_1,P_1]$.
Thus $\zeta [u_2,h_1,P_1]$ is satisfiable if and only if
$f_2 = \zeta[h_2,h_1,P_1]$ is satisfiable and 
$\zeta [ \zeta[u_3,h_1,P_1 ], f_2, P]$ is satisfiable
for a projection $P \in {\cal P}_{f_2,f_1}$. 
But since $f_2 \leq h_2$, it follows from Lemma \ref{lemma:Cofactors} that $P_2 \in {\cal P}_{f_2,f_1}$ and $P$ could be chosen to be $P_2$.
Continuing in this fashion
gives us the result.  The conclusion that the solution set of $f_k(x)=1$ is
the same as the solution set of $f(x)=1$ follows
from Corollary \ref{cor:satassign}. 
\epf

This result leads to a parallel algorithm
to solve the satisfiability problem.   This algorithm is described in Section \ref{sec:Algo}.

\section{An algorithm for SAT using projective co-factors} \label{sec:Algo}

From Theorem \ref{thm:ProjectiveDecomposition}, we obtain the algorithm
shown in Figure \ref{fig:Algo}.   The algorithm starts with a function defined
as a product $C_1.C_2. \ldots C_n$. 
At each step of the algorithm, we obtain
a set of potentially simple active functions out of which we select one
and check its satisfiability. In going from one step to the next, we form
projective cofactors of the remaining elements
in the set with the selected 
function (these can be performed
in parallel).  The algorithm stops
when the set of elements has only
one element whose satisfiability needs
to be checked.

\begin{figure} 
\begin{algorithmic}[H]
\State Given $f = h_1.h_2. \ldots h_k$, return SAT status, and assignment $A$
\State status = true, $A \leftarrow$ empty
\State Set $w_1 \leftarrow h_1, w_2 \leftarrow h2, \ldots w_k \leftarrow h_k$
\For {$i=1$; $i \leq k$; $i=i+1$} 
   \State $f_i \leftarrow w_i$
   \If {$f_i$ is not SAT} 
   \State status = false
   \State break
   \EndIf
   \If {$ i = k$} 
   \State break 
   \EndIf
   \State select projection $P_i \in {\cal P}_{f_i, h_{i+1}}$
   \For {$j=i+1$; $j \leq k$; $j=j+1$} 
      \State $w_j \leftarrow \zeta[w_j, f_i, P_i]$
   \EndFor
\EndFor
\If {status} 
   \State $A \leftarrow$ SAT-assignment for $f_k$
\EndIf
\State return (status, $A$)
\end{algorithmic}
\caption{A simple parallel algorithm for SAT based on projective decomposition}
\label{fig:Algo}
\end{figure}

We observe the following:
\begin{itemize}
\item The inner loop can be parallelized.
\item Satisfiability checking is performed only on the functions
$f_i$. This step can be very efficient if each $f_i$  depends on a small set of variables.
\item The main reduction step is the computation of
$\zeta [w_j, f_i, P_i]$, which can be efficient if $w_j$
and $f_i$ depend on a small number of variables.  Note that if
each of the $h_i's$ is simple, then finding an element in the
off-set of an $h_i$ is easy, and thus calculating a projection into
$h_i$ is also easy.
\item The projections $P_1$, $P_2$, \ldots $P_{k-1}$ can
be chosen arbitrarily.  In practice, the choice of projections
may have an impact on the effort involved in 
computing $\zeta[w_j,f_i,P_i]$.
\end{itemize}

\subsection{Choosing a projection $P$ onto a function $g$, and the calculation of $\zeta[f,g,P]$}

The critical step in the algorithm described in Figure \ref{fig:Algo} is
the computation of $\zeta[h,g,P]$ given functions $h,g$ and 
a projection $P \in {\cal P}_{g,h}$.  In general, $h$ and $g$ are
available as Boolean formulas, and we are interested in obtaining
a formula for $\zeta[h,g,P]$.
There are many possible ways of choosing the projection $P \in {\cal P}_{g,h}$. 
We illustrate a simple technique to find a projection 
and express it as a set of boolean formulas. Once the projection
is available as a set of boolean formulas, the actual calculation
of a formula for $\zeta[h,g,P]$ can then be carried out easily.

Consider the case that $g$ is a clause, and $h$
is available as a product of clauses.  Suppose that
\begin{displaymath}
g = x + y + z
\end{displaymath}
Now the projection must take the point $x=0,y=0,z=0$
and map it to some element in the off-set of $h$.  If $h$ is
available as a product of sums, then such an element can be
found easily.
We assume that an element of the
off-set of $h$ is known.  Let this element be $(u_1,u_2, \ldots u_n)$.
Note that if $h$ does not depend on variable $x_i$, then
we may choose $u_i = x_i$.
Then a projection $\in {\cal P}_{g,h}$ can be determined as follows: map
the point $(x_1,x_2, \ldots, x_n)$ to itself when it is in the
on-set of $g$, and onto $(u_1,u_2, \ldots u_n)$ when it
is not in the on-set of $g$.  The projection then becomes
\begin{eqnarray*}
x_1 & \rightarrow & g.x_1 + g'.u_1 \\
x_2 & \rightarrow & g.x_2 + g'.u_2\\
\ldots \\
x_n & \rightarrow & g.x_n + g'.u_n
\end{eqnarray*}
This can be simplified further.  For each $x_i$
which does not appear in $h$, we set $u_i = x_i$.
If, for some $q$, $u_q=0$, then $g.x_q + g'.u_q$  can be
simplified to $g.x_q$, and when $u_q = 1$, then
$g.x_q + g'.u_q$ can be simplified to $g' + x_q$.   Thus,
this projection takes
\begin{displaymath}
x_i \rightarrow \left\{ \begin{array}{ll}
x_i & {\rm when} \ h \ {\rm does \ not \ depend \ on \ } x_i \\
g.x_i & {\rm when} \ u_i=0 \\
g' + x_i & {\rm when} \ u_i = 1
\end{array}
\right.
\end{displaymath}

If $g$ itself can be written as a product $g_1.g_2.\ldots g_m$, then
we can find a projection in ${\cal P}_{g,h}$ by using Lemma
\ref{lemma:ProjComp}.  Choose $P_i \in {\cal P}_{g_i,h}$ for
$i=1,2, \ldots m$, and set $P$ to be the composition
\begin{displaymath}
(P_1 \circ P_2 \circ \ldots P_m)
\end{displaymath}

Instead of selecting a single element in $h_{OFF}$ to 
project all of $g_{OFF}$, a general procedure would be
to select, for each element $x$ such that $g(x) = 0$,
an element $u_x$ such that $h(u_x)=0$.  The resulting
Boolean functions would then need to be simplified to
obtain formulas describing the projection.  Finding
the projection with the simplest representation 
formula is something that needs to be studied further.

\section{Some Examples}

We will illustrate the use of 
Theorem \ref{thm:ProjectiveDecomposition} and the algorithm
shown in Figure \ref{fig:Algo} on two simple examples.  

\subsection{A satisfiable function }

Consider $f$ defined by 
\begin{displaymath}
f=C_{1}C_{2}C_{3}=(x'+y+w)(y'+z+w')(x+z+w')
\end{displaymath}
This function is satisfiable.    We
use Theorem \ref{thm:ProjectiveDecomposition} to confirm
that this is so.
\begin{itemize}
\item Let $f_1 = C_1$.  We choose the point
$y=1, z=0, w=1$ in the off-set of $C_2$ to 
construct the projection of $f_1$ into $C_2$.
The projection $P_1$ can then be worked out to be
\begin{eqnarray*}
x & \rightarrow & x \\
y & \rightarrow & y + C_1' = y + xw' \\
z & \rightarrow & z.C_1 = z.(x' + y + w)\\
w & \rightarrow & w + C_1' = w + xy' 
\end{eqnarray*}
\item We compute 
\begin{eqnarray*}
f_2 & = & \zeta[C_2,f_1,P_1] \\
    & = & (y'x'  + y'w + zx' + zy + zw + w'x' + w'y) 
\end{eqnarray*}
Similarly,
\begin{eqnarray*}
\zeta[C_3,f_1,P_1]  & =  & C_3
\end{eqnarray*}
\item For the projection $P_2$ of $f_2$ onto $C_3$, we choose
the point $x=0$, $z=0$, $w=1$ in the OFF-set $C_3$ and fix $P_2$
as:
\begin{eqnarray*}
x & \rightarrow & x.f_2 = xy'w + xyw' + xwz \\
y & \rightarrow & y \\
z & \rightarrow & z.f_2 = zx' + zy + zw \\
w & \rightarrow & w  + f_2' = (w+x).(w+y).(x' + y' + w).(x + z' + w).(x + y + w) 
\end{eqnarray*}
\item Using $P_2$, we obtain 
\begin{eqnarray*}
f_3 & = & \zeta[\zeta[C_3,f_1,P_1], f_2, P_2] \\
    & = & \zeta[C_3,f_2,P_2] \\
    & = & xy'w + xyw' + zx' + zy + zw + w'x' + w'y'
\end{eqnarray*}. 
\end{itemize}

Thus, by Theorem \ref{thm:ProjectiveDecomposition}, we 
observe that since $f_3$ is satisfiable,
$f$ is satisfiable.   Further, note that the solutions of
$f_3 = 1$ are exactly the solutions of $f = 1$.

\subsection{A non-satisfiable example}

Let
\begin{displaymath}
f = C_1 C_2  C_3 C_4 = (x + y)(x + y')(x' + y)(x' + y')
\end{displaymath}
In this case the function $f$ is not satisfiable.
Let us proceed with the algorithm implied by Theorem \ref{thm:ProjectiveDecomposition} to confirm this.
\begin{itemize}
\item Let $f_1 = C_1 = (x+y)$.  Fix the point $x=0$, $y=1$ in
the off-set of $C_2$ so that $P_1 \in {\cal P}_{C_1,C_2}$ is 
the projection:
\begin{eqnarray*}
x & \rightarrow &  x \\
y & \rightarrow &  x' + y
\end{eqnarray*}
For this projection we find
\begin{eqnarray*}
f_2 & = & \zeta[C_2,f_1,P1]  =   x  \\
w_{31} & = & \zeta[C_3,f_1,P_1]  =  x' + y \\
w_{41} & = & \zeta[C_4,f_1,P_1]   =  x' + y'
\end{eqnarray*}
\item 
Choose (using the point $x=1$, $y=0$ in the off-set of $C_3$) $P_2 \in {\cal P}_{f_2,C_3}$ to be the projection:
\begin{eqnarray*}
x & \rightarrow & 1 \\
y & \rightarrow & x.y
\end{eqnarray*}
Using this, we find 
\begin{eqnarray*}
f_3 & = & \zeta[w_31,f_2,P_2]  =  x.y \\
w_{42} & = & \zeta[w_{41},f_2,P_2]  =   x' + y' 
\end{eqnarray*}
\item Choose the element $x=1$, $y=1$ in
the offset of $C_4$ to obtain $P_3 \in {\cal P}_{f_3,C_4}$ as:
\begin{eqnarray*}
x  & \rightarrow & 1\\
y  & \rightarrow & 1
\end{eqnarray*}
We then find that
\begin{eqnarray*}
f_4 & =  & \zeta[w_{42}, f_3,P_3]\\
    & = & 0 
\end{eqnarray*}
\end{itemize}
Thus, by Theorem \ref{thm:ProjectiveDecomposition}, the
function $f$ is not satisfiable.

\section{Conclusions}

Given a Boolean function $f(x_1,x_2, \ldots, x_n)$ of $n$ variables, the satisfiability question asks if there is a 
point at which $f$ evaluates to $1$. An analogous consistency problem is to determine whether $f=0$ has a solution.   Decompositions of $f$ can help in solving such
problems.

If $f$ satisfies
\begin{displaymath}
f \leq g_1 + g_2 + \ldots g_m
\end{displaymath}
where $g_1,g_2, \ldots g_m$ are non-zero functions, then $f$ can be written as
\begin{displaymath}
f\ = \ \alpha_1.g_1 \ + \ \alpha_2.g_2 \ + \ \ldots \alpha_m . g_m
\end{displaymath}
where $\alpha_i$ is an arbitray Boolean function in the interval $[f.g_i, f+g_i']$ \cite{sude}.
Such an $\alpha_i$ is termed a co-factor of $g_i$ in $f$.
For this decomposition, $f$ is satisfiable if and only if $\alpha_i.g_i$ is satisfiable
for some $i \in \{1,2,\ldots, m\}$.   Thus, the original satisfiability problem can be
solved as $m$ parallel and independent problems.  If the functions $\alpha_i.g_i$
are simpler than $f$, there can be substantial reduction in computational
effort.

We have introduced the notion of a projective co-factor: 
If $g$ and $h$ are two Boolean functions ($h \neq 1$), then we define
a projection $P_{g,h}: B_0^n \rightarrow B_0^n$ which maps the ON-set of $g$
to itself and the OFF-set of $g$ into the OFF-set of $h$.
With respect to such a projection, we can define, for any
function $f$, a projective co-factor of $g$ in $f$ (denoted by $\zeta[f,g,P_{g,h}]$ which
maps ${\bf x}$ to $f(P_{g,h}({\bf x}))$.  We have shown that if $f = g.h$ with
$h \neq 1$, then $f$ is satisfiable if and only if $g$, and $\zeta[h,g,P_{g,h}]$ are 
{\em independently} satisfiable for any projection $P_{g,h}$ chosen as above.  
This result enables us to check the satisfiability of $f$ by separately (hence, in parallel) checking the satisfiability of two potentially simpler functions $g$ and $\zeta[h,g,P_g]$.  

From the projective decomposition result, we obtain a new  easily parallelizable transformational algorithm for the solution of a general satisfiability problem:
check the satisfiability of $f = g_1.g_2. \ldots g_m$ where each $g_i$ is a clause 
(or in the case of XOR-sat, an XOR-clause). The decomposition property can in
principle be used to devise a variety of decomposition schemes for solving
the satisfiability problem. Further algorithmic development and heuristics need to be investigated in order to determine
the practical feasibility of this approach.


\begin{thebibliography}{xxxx}
\bibitem{bool} George Boole. An Investigation of the Laws of thought. Walton, London, 1854.
\bibitem {knut} D.\ Knuth. Art of Computer Programming vol. 4.
\bibitem{brow} F.\ M.\ Brown. Boolean reasoning. The logic of Boolean equations. Dover, 2006.
\bibitem{ref:CoudertMadre} O.\ Coudert, J.\ Madre, A Unified Framework for the Formal Verification of 
Sequential Circuits, Proceedings of ICCAD, pp. 126-129, IEEE Press 1990.
\bibitem{rude} Sergiu Rudeanu. Boolean functions and equations. North Holland, Amsterdam, 1974.
\bibitem{hand} A.\ Biere, M.\ Heule, Hans van Maaren, T.\ Walsh (Eds). Handbook of Satisfiability. IOS Press, 2009.
\bibitem{crah} Yves Crama and Peter Hammer. Boolean functions. Theory, algorithms and applications. Encyclopedia of Mathematics and its applications, vol.142. Cambridge, 2011.
\bibitem{mezm} Marc Mezard and Andrea Montanari. Information, Physics and Computation. Oxford University Press, 2009.
\bibitem{hamw} Youssef Hammadi and C.\ M.\ Wintersteiger. Seven challenges in parallel SAT solving. Challenge paper AAAI 2012 Sub-Area spotlights track. Association of Advancement of Artificial Intelligence.
\bibitem{koha} Kohavi and Jha, Switching and automata theory, Cambridge 2008.
\bibitem{sul1} Virendra Sule, Generalization of Boole-Shannon expansion, consistency of Boolean equations and elimination by orthonormal expansion, arXiv.org/cs.CC/1306.2484v3, December 6, 2013.
\bibitem{sul2} Virendra Sule, An algorithm for Boolean satisfiability based on generalized orthonormal expansion, arXiv.org/cs.DS/1406.4712v3, July 16, 2014.
\bibitem{sude} M.P. Desai and Virendra Sule, Generalized cofactors and decomposition of Boolean satisfiability problems. arXiv.org/cs.DS/1412.2341v1, Dec 7, 2014.
\end{thebibliography}
\end{document}